\documentclass[12pt,twoside]{article}

\usepackage{a4wide}  
\usepackage{calc}
\usepackage{amsmath}
\usepackage{amsfonts,amssymb}
\usepackage[hyperindex]{hyperref} 
\usepackage{graphicx}

\newcommand{\lin}{l_{in}}
\newcommand{\lout}{l_{out}}
\newcommand{\Z}{\mathcal{Z}}
\newcommand{\D}{\mathcal{D}}
\newcommand{\genus}{\mathfrak{g}}

\newcommand{\mi}{\!-\!}
\newcommand{\equ}{\!=\!}
\newcommand{\pl}{\!+\!}

\newcommand{\klgl}{\!\leq\!}

\newcommand{\expec}[1]{\left<#1\right>}

\begin{document}

\title{
{\normalsize \hfill PITHA-06/04}\\
\vspace{-1.5cm}
{\normalsize \hfill ITP-UU-06/11}\\
\vspace{-1.5cm}
{\normalsize \hfill SPIN-06/09}\\
${}$\\ 
${}$\\ 
${}$\\ 
\Large \textbf{Nonperturbative sum over topologies in 2D Lorentzian quantum gravity\footnote{Talk given by S. Zohren at the Albert Einstein Century International Conference (Paris, July 18-22 2005).}} 
}
\author{{\large         R. Loll$^{1}$, W. Westra$^{1}$ and S. Zohren$^{1,2}$}\\[10pt]
        {\footnotesize \em 1) Institute for Theoretical Physics, Utrecht University}\\[-5pt]
        {\footnotesize \em Leuvenlaan 4, NL-3584 CE Utrecht, The Netherlands}\\
        {\footnotesize \em 2) Institut f\"ur Theoretische Physik E, RWTH-Aachen}\\[-5pt]
        {\footnotesize \em D-52056 Aachen, Germany\footnote{Since 01.10.2005 at Blackett Laboratory, Imperial College, London, SW7 2AZ, UK.}}\\[5pt]
        {\footnotesize \em E-Mail:}\\[-5pt]
        {\footnotesize  r.loll@phys.uu.nl, w.westra@phys.uu.nl and zohren@physik.rwth-aachen.de}
        }

\date{}
\maketitle

\begin{abstract}
The recent progress in the Causal Dynamical Triangulations (CDT) approach to quantum gravity indicates that gravitation is nonperturbatively renormalizable. We review some of the latest results in 1+1 and 3+1 dimensions with special emphasis on the 1+1 model. In particular we discuss a nonperturbative implementation of the sum over topologies in the gravitational path integral in 1+1 dimensions. The dynamics of this model shows that the presence of infinitesimal wormholes leads to a decrease in the effective cosmological constant. Similar ideas have been considered in the past by Coleman and others in the formal setting of 4D Euclidean path integrals. A remarkable property of the model is that in the continuum limit we obtain a finite space-time density of microscopic wormholes without assuming fundamental discreteness. This shows that one can in principle make sense out of a gravitational path integral including a sum over topologies, provided one imposes suitable kinematical restrictions on the state-space that preserve large scale causality.
\end{abstract}

\newpage


\section{Why quantum gravity?}

Quantum field theory has proven to be a marvelously successful way to describe three of the four fundamental forces of nature. For gravity however we do not have a well-defined predictive quantum field theoretic description yet, but we do have a very successful classical field theoretic description in the form of Einstein's general relativity. Since the other forces are well described by quantized field theories, it seems natural that there also exists a quantum theory of Einstein's general relativity.

Another reason for believing that such a theory of quantum gravity should exist is the fact that gravity is universal in the sense that it couples to all forms of energy. Hence the energy fluctuations at small distances due to Heisenberg's uncertainty relations induce also quantum fluctuations in the gravitational field. This leads to the prediction that space-time geometry has a highly non-trivial microstructure at extremely small scales proportional to the Planck length, $l_p\!=\!\sqrt{\hbar G_Nc^{-3}}\!\approx\!1.616\!\times\! 10^{-35}m$.

There are however obvious problems in constructing a quantum theory of general relativity. It has already been shown in the seventies by 't Hooft and Veltman that perturbative quantum gravity is non-renormalizable in four dimensions \cite{'tHooft:1974bx}. This does not mean that it is impossible to find a predictive theory of quantum general relativity. There are good indications that one can define a theory of quantum general relativity nonperturbatively \cite{Ambjorn:2005qt,Lauscher:2005xz}.

There are several nonperturbative approaches to quantum general relativity. Some of those attempts suggest that the ultraviolet divergences can be resolved by the existence of a minimal length scale, commonly expressed in terms of the Planck length $l_p$. A famous example is loop quantum gravity \cite{Ashtekar,Rovelli}; in this canonical quantization program the discrete spectra of area and volume operators are interpreted as evidence for fundamental discreteness. Other approaches, such as four-dimensional spin-foam models \cite{Perez:2003vx} or causal set theory \cite{Dowker}, postulate fundamental discreteness from the outset. Unfortunately, neither of these quantization programs has succeeded so far in recovering a sensible classical limit.

There are nonperturbative approaches which do not introduce a fundamental discreteness scale from the outset. One example is the exact renormalization group flow method for Euclidean quantum gravity in the continuum \cite{Lauscher:2005xz}. Another attempt is Causal Dynamical Triangulations (CDT), a covariant path integral formulation, in which Lorentzian quantum gravity is obtained as a continuum limit of a superposition of simplicial space-time geometries  \cite{Ambjorn:2005qt}.\\

In the following we summarize some of the recent successes of CDT in 3+1 dimensions, in particular, promising indications of recovering sensible classical behavior.

In addition, we give a short description of the CDT methods with special emphasis on the 1+1 dimensional case, since this model is exactly solvable and can be used to discuss fundamental issues that are also relevant for the higher dimensional realizations of CDT. One of these fundamental questions is whether the sum over topologies should be included in the path integral. As the main result we present a well-defined nonperturbative gravitational path integral including an explicit sum over topologies in the setting of CDT in 1+1 dimensions \cite{Loll:2003rn,Loll:2005dr}. A surprising feature of the model is that the presence of infinitesimal wormholes leads to a decrease in the effective cosmological constant, reminiscent of the suppression mechanism considered by Coleman and others in the four-dimensional Euclidean path integral \cite{Coleman:1988tj}.

\section{The CDT approach}

The CDT program is a quantization scheme for general relativity where no supersymmetry or ad hoc fundamental discreteness is assumed from the outset. The program is meant to give a rigorous nonperturbative definition of a path integral over all causal geometries\footnote{By this we mean the equivalence class of a Lorentzian metric modulo its diffeomorphisms.} weighted by the Einstein-Hilbert action

\begin{equation}
    \Z=  \int \D[g_{\mu\nu}] e^{i
    S_{\mathrm{EH}}[g_{\mu\nu}]}.
\end{equation}

From lattice QCD we know that discrete methods are a powerful tool to investigate nonperturbative effects in quantum field theory. In CDT one uses a specific discretization similar to Regge calculus where the geometry itself is encoded in a simplicial lattice. The advantage of using this type of discretization is that one is automatically working with gauge invariant degrees of freedom. There is no need to introduce coordinates in the construction \cite{Regge:1961px}. There is however a crucial difference between Regge calculus and dynamical triangulations. In the first the dynamics is encoded in the variation of the edge lengths whereas in dynamical triangulations the edge lengths are fixed but the dynamics is encoded in the gluing of the simplicial building blocks.

\begin{figure}
\begin{center}
\includegraphics[width=4in]{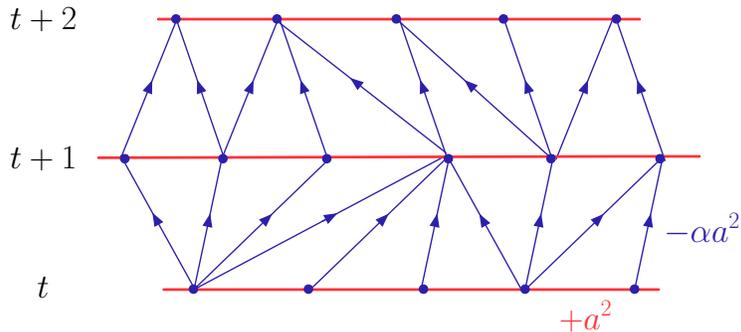}
\caption{A section of a sequence $[t,t+2]$ of two spacetime strips of a
triangulated two-dimensional spacetime contributing to the regularized path
integral.}\label{figure01}
\end{center}
\end{figure}

In the explicit construction of CDT the path integral over all causal geometries is written as a sum over all causal triangulations $T$ weighted by the Regge action (the simplicial analog of the Einstein-Hilbert action including a cosmological constant $\lambda$),
\begin{equation}\label{eq:Z}
\Z(\lambda,G_N)=\sum_{\text{causal T}}\frac{1}{C_T}e^{i S_{\mathrm{Regge}}},
\end{equation}
where $C_T$ is the discrete symmetry factor of the triangulation $T$. In contrast to previous attempts of dynamical triangulations the triangulations appearing in \textit{causal} dynamical triangulations have a definite foliated structure. In this type of triangulations each $(d\!-\!1)$-dimensional spatial slice is realized as an Euclidean triangulation whose simplicial building blocks have all squared edge lengths given by $l_s^2\!=\!a^2$. The successive spatial slices are connected by time-like edges of squared edge lengths $l_t^2\!=\!-\alpha a^2$ with $\alpha>0$, such that all building blocks in $T$ are $d$-simplices (see Fig. \ref{figure01} for an illustration in 1+1 dimensions). Here the parameter $a$ is a cut-off length that one takes to zero in order to obtain the continuum limit of the regularized path integral \eqref{eq:Z}. Note that in this limit the individual triangulations correspond to the individual histories of the path integral, which in general do not resemble smooth manifolds.

The foliated structure of the triangulations introduces a natural notion of discrete global time $t$ given by the label of the successive spatial slices. Note that one has to be careful in attaching a physical meaning to this global time as we discuss later on. The clear distinction between space-like edges and time-like edges enables us to define a Wick rotation on each causal triangulation by analytic continuation of $\alpha\!\mapsto\!-\alpha$. It is important to realize that the set of Euclidean triangulations one obtains after the Wick rotation is strictly smaller than the set of all Euclidean triangulations.

\subsection{Recent results in 3+1 dimensions}

Before describing the analytic results in 1+1 dimensions let us mention some of the recent exciting successes of CDT in 3+1 dimensions as a motivation for the CDT approach to quantum gravity  (see also \cite{Ambjorn:2005jj} for a general overview).

In absence of an analytic solution for the 3+1 dimensional model, one uses Monte Carlo simulations to obtain numerical results. A very important non-trivial test for every nonperturbative formulation of quantum gravity is whether it can reproduce a sensible classical limit at macroscopic scales.
The numerical results indicate that the scaling behavior of the spatial volume as a function of space-time volume is that of a four-dimensional universe at large scales, a first indication of sensible classical behavior \cite{Ambjorn:2004qm}. Moreover, after integrating out all dynamical variables apart from the spatial volume as a function of proper time, one can derive the scale factor whose dynamics is described by the simplest minisuperspace model used in quantum cosmology \cite{Ambjorn:2004pw}.

Having passed the first consistency checks regarding the \textit{macro}scopical structure of space-time it is very interesting what predictions one can make for the quantum nature of the \textit{micro}structure of space-time. One important observable which has been measured is the spectral dimension of space-time which is the dimension a diffusion process would feel on the space-time ensemble. Surprisingly, this quantity depends on the scale at which it is measured. More precisely, one observes a dimensional reduction from four at large scales to two at small scales within measurement accuracy \cite{Ambjorn:2005db}. This gives an indication that nonperturbative quantum gravity defined through CDT provides an effective ultraviolet cut-off through a dynamical dimensional reduction of space-time.

\subsection{Analytic results in 1+1 dimensions}

To better understand the methods used in CDT it is useful to look in more detail at the 1+1 dimensional model, since it is exactly solvable \cite{Ambjorn:1998xu}.

\begin{figure}
\begin{center}
\includegraphics[width=2.2in]{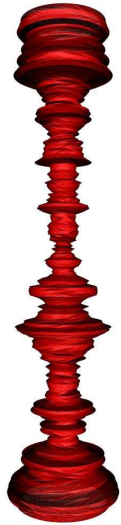}
\caption{A typical two-dimensional Lorentzian space-time. The compactified direction shows the spatial hypersurfaces of length $L$ and the vertical axis labels time $T$. Technically, the picture was generated by a Monte Carlo simulation, where a total volume of $N=18816$ triangles and a total time of $t=168$ steps was used. Further, initial and final boundary has been identified.}\label{figure03}
\end{center}
\end{figure}

Recall that the Einstein-Hilbert action in two dimensions is given by
\begin{equation}\label{eq:EH}
S_{\mathrm{EH}}[g]=\int_M d^2x \sqrt{|\det g|}\,\Lambda - 2\pi K \chi(M)
\end{equation}
where $K\equ G_N^{-1}$ is the inverse Newton's constant, $\chi(M)\equ 2\mi 2\genus$ the Euler characteristic of the manifold $M$ and $\genus$ the genus of $M$. Therefore, for fixed spatial topology, the curvature term in the action contributes just a constant phase factor to the path integral. If one allows for topology changes this term becomes important for the quantum dynamics as we will see in the next section. In this section we fix the topology of space-time to be $\mathbb{R}\!\times\!S^1$. The most natural thing to calculate is the propagator from an initial geometry of length $\lin$ to a final geometry of length $\lout$ in time $t$. Using the discrete analog of the Einstein-Hilbert action \eqref{eq:EH} one can write down the propagator after Wick rotation as the path integral \eqref{eq:Z} for fixed boundaries $\lin$ and $\lout$,
\begin{equation}\label{eq:G}
G_{\lambda}(l_{in},l_{out};t)=\sum_{\substack{\text{causal T}:\\ l_{in}\rightarrow l_{out}}}e^{-\,\lambda\,a^2\,N(T)},
\end{equation}
where $\lambda$ is the bare cosmological constant and $N(T)$ the number of triangles in the triangulation. In two dimensions the sum in \eqref{eq:G} can be evaluated and one obtains the continuum limit after a suitable choice of renormalization, yielding the continuum propagator
\begin{eqnarray}\label{eq:Gcont}
    G_\Lambda(L_{in},L_{out};T)=\sqrt{\frac{2\Lambda}{L_{in}L_{out}}}
    \frac{e^{-\sqrt{2 \Lambda}(L_{in}+L_{out})\coth(\sqrt{2\Lambda}T)}}{\sinh(\sqrt{2\Lambda}T)}
    I_1 \left(2 \frac{\sqrt{2 \Lambda L_{in}L_{out}}}{\sinh(\sqrt{2\Lambda}T)}\right),
\end{eqnarray}
where $I_1(x)$ denotes the modified Bessel function of the first kind and $\Lambda$, $L_i$ and $T$ are the continuum counterparts of $\lambda$, $l_i$ and $t$. Physically this solution can be interpreted as a fluctuating two-dimensional ``universe'' (Fig. \ref{figure03}), where the average spatial length and its fluctuations are determined by the cosmological constant, $\expec{L}\!\sim\!\expec{\Delta L}\!\sim\!1/\sqrt{\Lambda}$.

Remarkably, the continuum propagator \eqref{eq:Gcont} agrees with the result of the propagator obtained from a continuum calculation in the proper-time gauge of 1+1 dimensional pure gravity \cite{Nakayama:1993we}. This indicates that the above choice of global time is similar to the one used in the proper-time gauge. Another interesting question is whether the result obtained in \eqref{eq:Gcont} is independent of the choice of foliation. There are good indications that due to the broad universality class of this model one also obtains the same dynamics \eqref{eq:Gcont} for different choices of time slicing \cite{Arnsdorf:2001wh}.

\section{Including topology changes}

A recurring question in the history of quantum gravity approaches is whether one should allow for topology changes of space-time. In terms of path integrals this translates into whether one should include the sum over topologies in the path integral
\begin{equation}\label{contPI}
    \Z= \sum_{\mathrm{topol.}} \int \D[g_{\mu\nu}] e^{i
    S_{\mathrm{EH}}[g_{\mu\nu}]}.
\end{equation}

There have been several attempts to solve the path integral \eqref{contPI} in the setting of Euclidean quantum gravity. The big problem however that becomes apparent even in the simplest case of two dimensions is that the full sum over topologies cannot be uniquely defined nonperturbatively, since the sum over genera is badly divergent. One of the main differences between this approach and the one we propose here is that we restrict the class of topology changes by means of imposing an (almost everywhere) causal structure. In the following we show that this restriction on the topology changes leads to a better defined path integral and we introduce a model with infinitesimal wormholes where one can perform the sum over topologies explicitly \cite{Loll:2003rn,Loll:2003yu,Loll:2005dr}.

\subsection{Construction of the model}

\begin{figure}
\begin{center}
\includegraphics[width=3in]{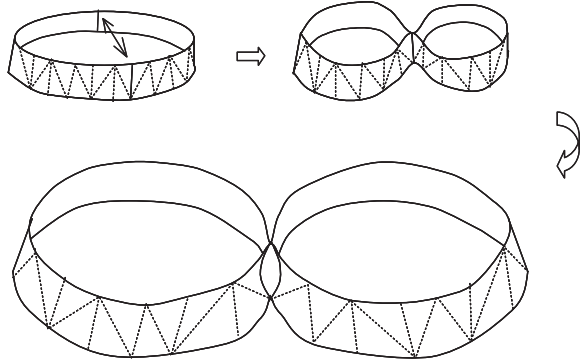}
\caption{An infinitesimal wormhole is constructed in a space-time strip of topology $[0,1]\times S^1$ by identifying two of the time-like edges and cutting open the
geometry along this edge.}\label{figure02}
\end{center}
\end{figure}

We define the sum over topologies in \eqref{contPI} by performing
surgery moves directly on the triangulations to obtain regularized
versions of higher-genus manifolds \cite{Loll:2003rn,Loll:2003yu}.
For the construction of these moves let us concentrate on a single
space-time strip of topology $[0,1]\times S^1$ and height $\Delta
t =1$ as illustrated in Fig. \ref{figure02}. The infinitesimal
wormholes can be constructed by identifying two of the time-like
edges and subsequently cutting open the geometry along this edge.
By applying this procedure repeatedly and obeying certain
causality constraints \cite{Loll:2003rn,Loll:2003yu}, more and
more wormholes can be created.

\subsection{Solution of the model}

To obtain the dynamics of the model it is sufficient to analyze the one-step propagator. Including the topological term in the action and performing the Wick rotation gives
\begin{equation}\label{eq:topotrans:sumovertop}
G_{\lambda,\kappa}(l_{in},l_{out};t=1)= \sum_{\substack{\text{causal T}: \\ l_{in}\rightarrow l_{out}}}e^{-\lambda\,a^2\,N-2\kappa\, \genus (T)},
\end{equation}
where the sum is taken over all possible triangulations $T$ of
height $t\!=\!1$ with fixed initial boundary $l_{in}$ and final
boundary $l_{out}$, but arbitrary genus $0\klgl\genus (T)\klgl
\left[N/2\right]$ (here $N\equ l_{in}\pl l_{out}$ is the number of
triangles in $T$, which coincides with the number of time-like
edges). Further, $\lambda$ is the bare cosmological constant and
$\kappa$ is the bare inverse Newton's constant. The sum over all
possible triangulations with arbitrary genus can be
performed unambiguously and one can obtain the continuum limit after a suitable
choice of renormalization and double scaling limit for $\lambda$
and $\kappa$, yielding the continuum propagator \cite{Loll:2005dr}
\begin{eqnarray}\label{eq:Gconttop}
    G_{\Lambda,G_N}(L_{in},L_{out};T)=\sqrt{\frac{2\Lambda_{eff}}{L_{in}L_{out}}}
    \frac{e^{-\sqrt{2 \Lambda_{eff}}(L_{in}+L_{out})\coth(\sqrt{2\Lambda_{eff}}T)}}{\sinh(\sqrt{2\Lambda_{eff}}T)}
    I_1 \left(2 \frac{\sqrt{2 \Lambda_{eff} L_{in}L_{out}}}{\sinh(\sqrt{2\Lambda_{eff}}T)}\right),
\end{eqnarray}
where $\Lambda_{eff}$ is given in terms of the renormalized cosmological and Newton's constant as $\Lambda_{eff}=\Lambda(1-e^{-4\pi/G_N})$.

In addition to the known geometrical observables this model possesses a new type of topological observable, namely, the density of wormholes, which can be calculated to give a finite expression
\begin{equation}
\label{densityonG}
n=\frac{\expec{N_\genus}}{\expec{V}}=
\frac{1}{e^\frac{4\pi}{G_N}-1} \,\Lambda.
\end{equation}

\begin{figure}
\begin{center}
\includegraphics[width=4in]{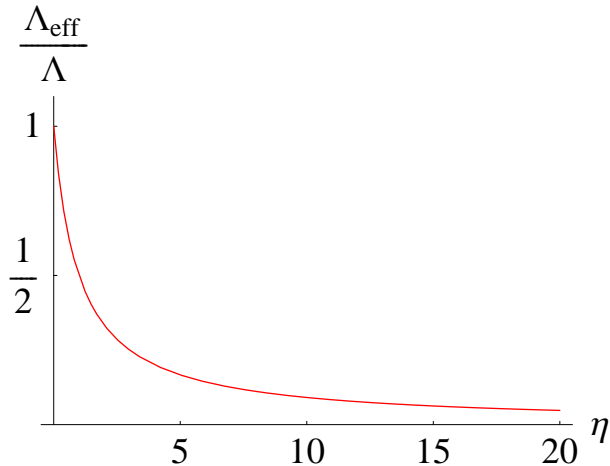}
\caption{The ``effective'' cosmological constant $\Lambda_{eff}$ in units of $\Lambda$ as a function of the density of infinitesimal wormholes in units of $\Lambda$, $\eta\equ n/\Lambda$.}\label{figure04}
\end{center}
\end{figure}

It is useful to reinterpret the physical system
described by \eqref{eq:Gconttop} in terms of its physical
quantities, namely the cosmological constant $\Lambda$ and the
density of wormholes in units of $\Lambda$, i.e.
$\eta\equ\frac{n}{\Lambda}$. These two quantities can be seen to
set the physical scales of the system. Whereas in the case without
topology changes (\ref{eq:Gcont}) there was only one scale, namely,
the cosmological scale $\expec{L}\!\sim\! 1/\sqrt{\Lambda}$, in
the case with topology changes there is in addition the relative
scale $\eta$ between cosmological and topological fluctuations.
Both together define the effective fluctuations in length through
$\expec{L}\!\sim\! 1/\sqrt{\Lambda_{eff}}$, where we can now write
$\Lambda_{eff}\equ\Lambda/(1\pl\eta)$. The behavior of
$\Lambda_{eff}$ as a function of $\eta$ is shown in Fig.
\ref{figure04}. One can observe that the presence of wormholes in
spacetime leads to a decrease in the ``effective'' cosmological
constant $\Lambda_{eff}$.

This connects nicely to former attempts to derive a mechanism, the
so-called Coleman's mechanism, to explain the smallness of the
cosmological constant in a formal Euclidean path integral formulation
of four-dimensional quantum gravity in the continuum with the
presence of infinitesimal wormholes \cite{Coleman:1988tj}. The
wormholes considered in those models resemble those of our toy
model in that both are non-local identifications of the spacetime
geometry of infinitesimal size. The counting of the wormholes
considered here is of course different since we are working in a
genuinely causal and background independent setup which enables us
to actually perform the sum over topologies explicitly, without
assuming any information on a background manifold.

\section{Conclusion}

In this paper we briefly reviewed some of the recent successes of
the CDT approach to quantum gravity. The existence of a
semiclassical ground state in the computer simulations of the
$3+1$ dimensional CDT model is perhaps the most tantalizing hint
that gravitation can be quantized nonperturbatively. The main
message we would like to convey in this contribution is that the
CDT method is also ideally suited for investigating nonperturbative
topology change. Here we have shown that one can analytically sum
over all genera in a $1+1$ dimensonal gravitational path integral
provided suitable kinematical causality restrictions are imposed.
A surprising effect of including these topology changes is that
they tend to lower the value of the cosmological constant. It would be interesting to see whether the causally restricted topology changes would also lead to similar results in higher dimensional implementations. We will return to this issue in future work.


\section*{Acknowledgements}
S.Z. thanks J. Jers\'ak for valuable comments and acknowledges
support of the Dr. Carl Duisberg-Foundation
and a Theoretical Physics Utrecht Scholarship.
R.L. acknowledges support by the Netherlands Organisation for
Scientific Research (NWO) under their VICI program. This work was partly funded through the Marie Curie Research and Training Network ENRAGE (MRTN-CT-2004-005616).

\end{document}